# Tri-layer SiN-on-Si 8×8 Optical Switches with Thermo-optic and Electro-optic Actuators

Bohao Sun, Chunhui Yao, Tongyun Li, Ziyao Zhang, Peng Bao, Minjia Chen, Alan Yilun Yuan, Chenxi Tan, Zhitian Shi, Adrian Wonfor, Seb Savory, Keren Bergman, Richard Penty, Qixiang Cheng

*Abstract*—We present two spatial-multiplexed switch-and-select (S&S) 8×8 optical switches incorporating a tri-layer SiN-on-Si platform, one equipped with thermo-optic (T-O) and the other electro-optic (E-O) switching elements. To the best of our knowledge, the electro-optic switch fabric is the first-of-its-kind device assembled in such a multi-layer platform. The shuffle between the multiplexer and demultiplexer array is established via a tri-layer Si-SiN-SiN structure, creating a three-dimensional crossing-free photonic shuffle network. At the same time, the implementation of the S&S topology can effectively suppress the first-order crosstalk. The measured on-chip losses for the T-O switch range from 2.1 to 11.5 dB, with a 5.2 dB average, while the E-O device exhibits losses between 8.7 to 19.6 dB, with a 15.1 dB average. Both switches demonstrate ultra-low crosstalk, with measured ranges of 38.9 to 50.8 dB and 42.8 to 51.9 dB, for the T-O and E-O devices respectively. The switching times are 17.6 μs for the T-O switch and 5.9 ns with the E-O actuated one. These performance metrics highlight the potential of these switches for next-generation data center applications.

## I. INTRODUCTION

THE surge of data-intensive applications, such as cloud computing and artificial intelligence, has increasingly driven modern data centers to rely on interconnect networks with enhanced data capacity [1], [2]. Optical switching technologies offer large bandwidth, low latency, and high power efficiency compared to traditional electronic switches, positioning them to play a crucial role in meeting the evolving demands of future data communication networks [3]. One of the most promising platforms for optical switches is silicon photonics, which leverages the mature Complementary Metal-Oxide-Semiconductor (CMOS) process, providing a cost-effective solution with high fabrication precision. Technically, optical switching can be achieved through various approaches, including liquid crystals on Silicon (LCOS) [4], micro electromechanical systems (MEMSs) [5], Mach–Zehnder interferometers (MZIs) [6], [7], micro-ring/disk resonators [8-11], semiconductor optical amplifiers (SOAs) [12], and so on. Among these, resonator-based switches, particularly micro-rings (MRRs) and microdisks, stand out for their compactness and ability to efficiently manipulate light at the wavelength scale.

In silicon photonic switch fabrics, phase shifting is typically achieved through two primary mechanisms: T-O and E-O modulation [11]. T-O switches, which manipulate the refractive index through temperature changes, generally offer lower optical loss compared to E-O devices that rely on the free-carrier dispersion effect to alter the refractive index.[13], [14], Yet, E-O switches provide a key advantage of switching speed, with switching times that can reach the nanosecond scale. Both T-O and E-O mechanisms are often applied to MRRs or microdisks to enable efficient optical switching. MRRs operate with bent waveguide modes, while microdisks harness whispering gallery modes [15]. As a result, microdisks tend to be more compact, with a larger free-spectral range (FSR) and better fabrication tolerance [15], [16].

For both types of micro-resonators, the S&S architecture is recognized as a suitable candidate for constructing an optical switch fabric, owing to its immunity to first-order crosstalk and strictly non-blocking connectivity. However, as the number of ports in the switch system increases, a key challenge arises: the large number of waveguide crossings in the central shuffle of such an S&S network significantly limits its scalability [8], because the number of waveguides required scales quadratically with the number of ports. This issue is particularly pronounced in traditional two-dimensional (2D) photonic platforms with in-plane waveguide crossings. To tackle this challenge, additional SiN layers have been introduced to form crossing-free tri-layer shuffles [17]. Table 1 presents a performance comparison of switches with S&S topology across different platforms, port counts, and switch elements. The results highlight the advantage of three-dimensional (3D) crossings in achieving ultralow crosstalk. To the best of our

Manuscript received XXXX XX, XXXX; revised XXXX XX, XXXX; accepted XXXX XX, XXXX. This work was supported in part by the U.S. ARPA-E under ENLITENED Grant DE-AR00084, the UKRI-EPSRC by project QUDOS (EP/T028475/1), and the European Union's Horizon Europe Research and Innovation Program project PUNCH (101070560) and project INSPIRE (101017088). (Corresponding author: Qixiang Cheng)

Bohao Sun, Chunhui Yao, Tongyun Li, Ziyao Zhang, Peng Bao, Minjia Chen, Alan Yilun Yuan, Chenxi Tan, Zhitian Shi, Adrian Wonfor, Seb Savory, Richard Penty and Qixiang Cheng are with the Electrical Engineering Division, Department of Engineering, University of Cambridge, Cambridge CB3 0FA, UK (e-mail: bs665@cam.ac.uk, cy327@cam.ac.uk, tl299@cam.ac.uk, zz447@cam.ac.uk, pb771@cam.ac.uk, mc2243@cam.ac.uk, ayy21@cam.ac.uk, ct633@cam.ac.uk, zs411@cam.ac.uk, aw300@cam.ac.uk, sjs1001@cam.ac.uk, rvp11@cam.ac.uk, and qc223@cam.ac.uk). Qixiang Cheng and Chunhui Yao are also with GlitterinTech Limited, Xuzhou, China.

Keren Bergman is with Center for Integrated Science and Engineering, Dept of Electrical Engineering, Columbia University, City of New York NY 10027, United States (e-mail: bergman@ee.columbia.edu).

TABLE I
PHOTONIC INTEGRATED SWITCHES IN THE S&S TOPOLOGY

| Work | Platform | Port Count | Topology and Switch Element | Loss (dB) | Crosstalk (dB) | Switching Speed |
|---|---|---|---|---|---|---|
| Ref. [18] | Si | 4 × 4 | S&S MRR | 4.5 – 19 | -28.3 – -46.7 | N/A |
| Ref. [8] | Si-SiN | 4 × 4 | S&S MRR | 1.8 – 20.4 | -31.4 – -52.7 | 14.3 μs |
| Ref. [19] | Si | 8 × 8 | S&S MZI | 4 – 6.8 | -30 – -45 | 250 μs |
| Ref. [20] | Si-SiN-SiN | 32 × 32 | S&S MZI | 13.73 – 19.12 | -34.98 – -60.65 | 44.8 μs |
| This work (T-O) | Si-SiN-SiN | 8 × 8 | S&S MRR | 2.1 – 11.5 | -38.9 – -50.8 | 17.6 μs |
| This work (E-O) | Si-SiN-SiN | 8 × 8 | S&S Microdisk | 8.7 – 19.6 | -42.8 – -51.9 | 5.9 ns |

knowledge, in this paper, we present the first Si-SiN-SiN microresonator based E-O switch.

In this paper, we demonstrate two 8×8 ultralow-crosstalk Si-SiN-SiN optical switches in the S&S architecture, using T-O MRRs and E-O microdisks as switching elements, respectively. Each switch fabric is assembled using 16 spatial (de)multiplexers, incorporating 128 switch elements in total. To the best of our knowledge, the E-O switch fabric represents the first-of-its-kind device implemented in such a multi-layer platform. The T-O and E-O switches show an on-chip insertion loss of 2.1 - 11.5 dB and 8.7 - 19.6 dB, with an average loss of 5.2 dB and 15.1 dB, respectively. These two switches also demonstrate ultralow-crosstalk of 38.9-50.8 dB (T-O) and 42.8-51.9 dB (E-O). The 3 dB passbands are greater than 70 GHz and 86 GHz, with switching times measured at 17.6 μs and 5.9 ns, respectively. Additionally, data transmission at 25 Gb/s using an On-Off Keying (OOK) signal was evaluated through the E-O switch fabric, illustrating high-fidelity performance.

## II. OPTICAL SWITCH DESIGN AND PACKAGING

### A. Switch Topology

Both optical switches employ the S&S topology, a strictly non-blocking switch architecture that is able to create any input to output path permutation without the need to reset any already set up paths [8]. Figure 1(a), (b), and (c) show the structures of these two S&S topology 8×8 optical switches, with T-O MRRs and E-O microdisks based switching cells as examples. Figure 1(d) and (e) show the 3D structures of switching elements (MRR and second-order microdisk). In these designs, the function of switching and selecting is achieved via spatial (de)multiplexers, via either a 1×8 or 8×1 switch element bank. The total number of switch elements is 128.

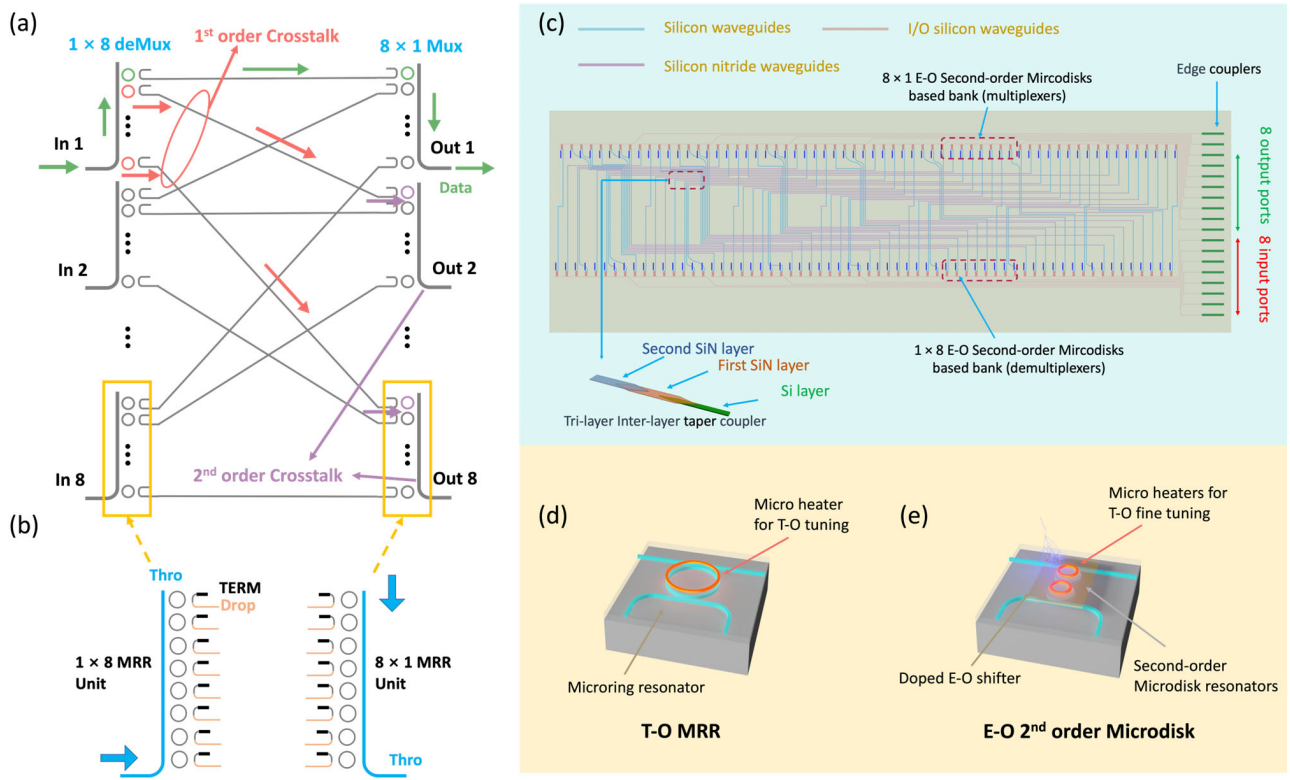

Fig. 1. (a) Schematic of 8×8 S&S MRR based optical switch. Green arrows describe the data paths, red and purple arrows represent first-order and second-order crosstalk respectively. (b) Schematics of the multiplexer and demultiplexer. (c) Diagram of the 8×8 S&S E-O switch showing the shuffle network, together with the Si-SiN-SiN inter-layer coupler. (d) 3D model of the T-O switching element. (e) 3D model of the E-O switching element.

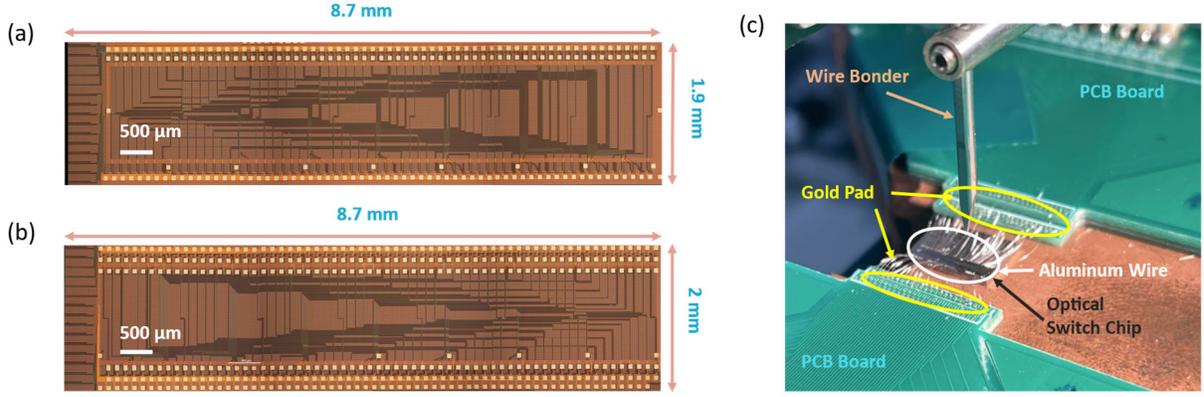

Fig. 2. (a) Microscope photo of the T-O switch. (b) Microscope photo of the E-O switch. (c) Photo of the wire-bonded optical switch chip.

The Si-SiN-SiN tri-layer central shuffle creates a waveguide crossing-free network to connect multiplexers and demultiplexers. In this work, a total of 784 in-plane waveguide crossings are avoided in each T-O and E-O 8×8 switch-and-select optical switch, with an average and maximum numbers of crossings avoided in each path being approximately 12 and 35, respectively. Based on the test structure measurements, the insertion loss and crosstalk for each Si-SiN 3D waveguide crossing are 0.012 dB and <-55 dB, respectively. Each 3D crossing features a more compact footprint of 1.5×0.4 μm$^2$ compared to Si single-layer crossings [21], leading to a total crossing footprint of only 470.4 μm$^2$ per switch.

Specifically, to establish the path from input port a to output port b (path a-b), the $b^{th}$ element in the $a^{th}$ bank would be connected to the $a^{th}$ element in the $b^{th}$ bank, with only the two elements in resonance (on-state). To minimize switching loss and crosstalk, the remaining elements are all set in a non-resonant state (off-state). The first-order crosstalk is suppressed by the off-state switch elements in the other banks. As a result, crosstalk leakage to other output ports is greatly diminished as both the input and output microdisks are off, each producing attenuation.

### B. Chip Packaging

Microscope photos of the T-O and E-O switches are shown in Figure 2 (a) and (b), respectively. The two chips are in-house packaged with wire bonding and customized printed circuit boards (PCBs) (Figure 2 (c)). The T-O switch requires a total of 130 pads, comprising 128 pads for the 64 T-O MRRs and 2 pads for ground connections. In contrast, the E-O switch requires 386 pads in total. This is because each microdisk switch cell consists of an E-O phase shifter and two T-O phase tuners, accounting for 384 pads dedicated to the 64 microdisk switch cells, along with 2 pads for ground connections. To accommodate this, we use customized PCBs placed on either side of the chips to provide sufficient electrical fan-out. Edge couplers are used to transfer the optical signal between the fiber array and the optical switch. In our experiment, a 6-axis optical stage and a 24-channel fiber array are used to couple light into the optical switch.

### C. Experimental Setup

Figure 3 illustrates the experimental setup for measuring both switches. Green arrows represent electrical paths, and yellow arrows indicate optical paths. A control program on a PC using

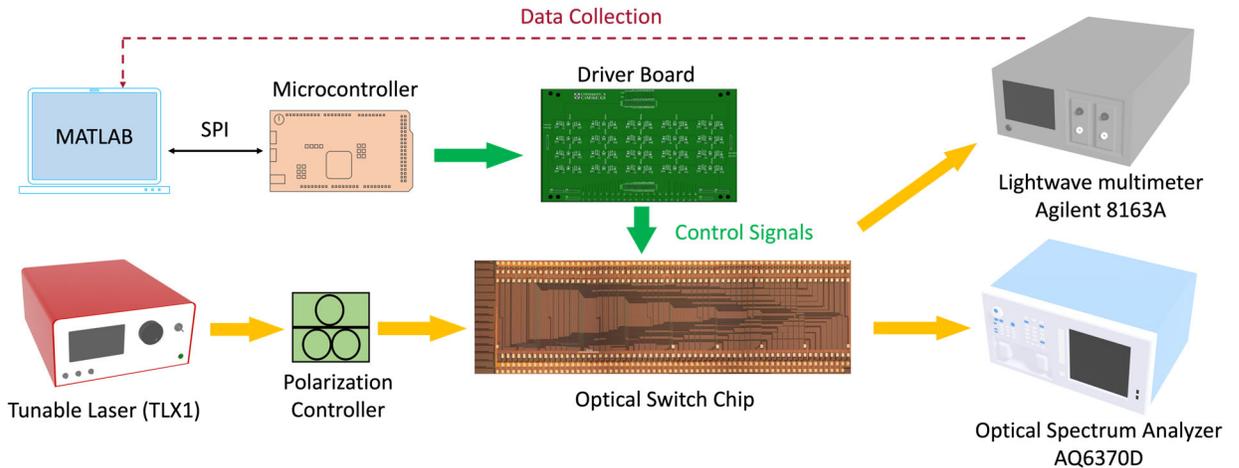

Fig. 3. Experimental setup and structure with an 8×8 optical switch photo.

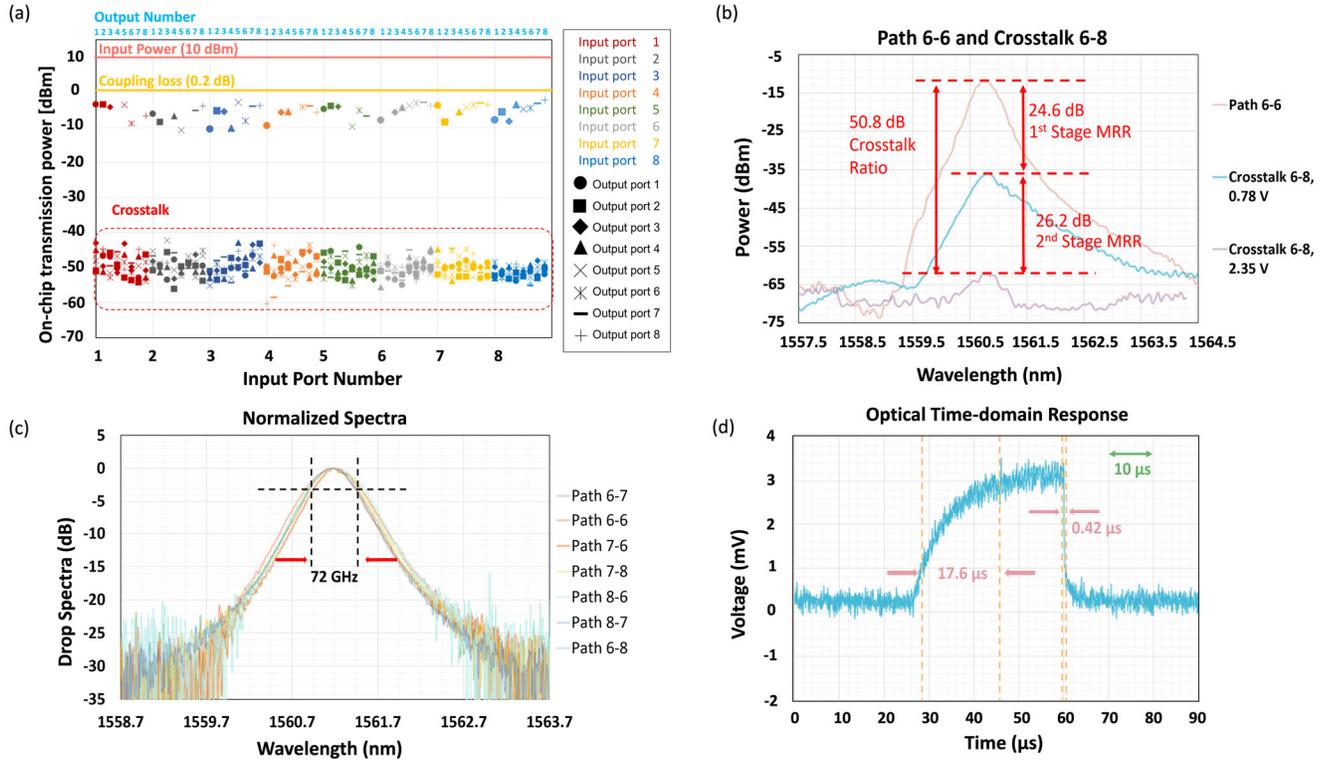

Fig. 4. (a) Measured optical power map of 8 × 8 MRRs based T-O switch. (b) Crosstalk ratio measurement: signal power at output port 6 and crosstalk at output port 8 with two MRR states. (c) Normalized transmission spectra for several representative optical paths. (d) Optical time-domain response of T-O switch.

MATLAB is used to generate switching control signals, incorporating a voltage-ring routing table. The PC interfaces with a microcontroller through the universal asynchronous receiver/transmitter (UART) port, which subsequently sends the control signals to 40-channel 16-bit digital-to-analog converter (DAC) boards through their serial peripheral interface (SPI) ports, enabling precise generation of chip control voltages (-4V to 8 V) with a resolution of 0.2 mV. A dedicated switch driver board is employed, which features an array of ADA4625 operational amplifiers with an 11.5 dB gain at the DAC output to meet the voltage requirements of the MRRs (0-1.5 V) and those of the microdisk E-O switching elements (0-1.3 V). The control signals are then applied to the heaters to execute the desired switching operation.

A C-band tunable laser and lightwave multimeter, connected to the PC, facilitate fully automated calibration of the switch and determine the optimum control voltage for each route. After the calibration, a voltage look-up table for on- and off-states of the 128 rings can be created and stored in the control program.

### III. T-O Switch Structure and Performance

#### A. Photonic Device Design and Fabrication

The T-O switch chip was taped out for an AIM Photonics Si-SiN-SiN multi-project wafer (MPW) run with a footprint of 1.9×8.7 mm$^2$. The central shuffle network is based on a Si-SiN-SiN tri-layer structure with interlayer couplers and 3D waveguide crossings. The first SiN layer acts as an intermediate layer, facilitating efficient light transition from the base Si layer to the upper SiN layer. By introducing this intermediate layer, the gap between the Si waveguide and the top SiN waveguide can be effectively increased, and accordingly reduce the crossing loss and crosstalk [20].

#### B. Optical Power Map of the T-O Switch

Figure 4(a) shows the measured optical power map of the 8 × 8 T-O switch (coupling loss of 4.9 dB per facet), with an incident optical power of 10 dBm at a wavelength of 1561 nm. The red dashed region highlights the crosstalk leakage. Note that 10 out of 64 paths are unavailable in Fig. 4(a) due to fabrication defects and wire bonding damage. According to the optical power map, the on-chip optical loss and crosstalk ratio are measured in the range of 2.1 to 10.5 dB and -38.9 to -50.8 dB respectively. In some paths, such as path 8-8, there is no Si-SiN 3D waveguide crossing or interlayer coupler, and this path includes only two on-state MRRs. Thus path 8-8 represents the case with the lowest loss (2.1 dB). In contrast, path 1-1 also lacks waveguide crossings and interlayer couplers, but it includes 14 off-state MRRs and 2 on-state MRRs, resulting in higher optical loss (3.6 dB). Additionally, the propagation loss in the Si and SiN waveguides contributes to the varying overall optical loss for different paths. A tunable laser is employed to generate a continuous-wave signal at C-band wavelengths. The FSR of this T-O switch is measured to be around 25.6 nm.

#### C. Extinction Ratio and Crosstalk of the T-O Switch

The extinction ratio of each MRR is a significant parameter to evaluate the performance of the optical switch. All measured MRRs feature an on-off extinction ratio exceeding 25 dB. Figure 4(b) shows an example of the crosstalk leakage into output port 8 when routing optical signal via path 6-6. The red curve represents the data transmission of path 6-6, while the light blue and purple curves indicate the channel response when the 6th MRR at the 8th 8×1 MRR bus are respectively biased on and off. It can be seen from Figure 4(b) that the first and second-order crosstalk ratios are 24.6 dB and 26.2 dB individually, so the total crosstalk ratio is 50.8 dB between the output ports 6 and 8.

*D．Optical Passband, Switching Power, and Switching Time*

The normalized optical transmission spectra of representative optical paths are shown in Figure 4(c), with a 3 dB passband of approximately 72 GHz for the T-O MRRs-based switch. The on-state bias voltage of the MRRs ranges from 1.6 to 2.2 V, while the off-state bias voltage is between 0 and 0.4 V. The total power consumption per path, considering both on-state and off-state tuning, is around 35 mW.

A function generator producing a 15 kHz square wave signal with a 50% duty cycle is used to modulate specific MRRs to establish the corresponding path for on/off switching. The time-domain response of the output signal is recorded on an oscilloscope. Figure 4(d) indicates that the switching rise and fall times (10 %-90 %) are 17.6 μs and 0.42 μs, respectively. The observed difference in rise and fall times is because of the thermal conductivity of the chip [22]. This phenomenon is also observed in [8]. The thermal propagation speed limits the switching rise time, at the same time, the shorter switching fall time is caused by the rapid heat dissipation. One solution to balance the switching time is to introduce thermal isolation trenches, which can equalize the heating and cooling dynamics during the chip operation [23].

## IV. E-O SWITCH STRUCTURE AND PERFORMANCE

*A. Photonic Device Design of the E-O Switch*

E-O microdisk switch occupies a total footprint of 2×8.7 mm$^2$. The central shuffle network is built on a Si-SiN-SiN platform, similar to that of the T-O switch, enabling each input to connect to each output without waveguide crossings. Establishing each optical path necessitates only the tuning of a single pair of second-order microdisks to resonance, while the other components in the corresponding bus banks are not activated, i.e. in their off-resonance state, minimizing the channel crosstalk value. Each microdisk switch cell is equipped with an E-O phase shifter and two thermo-optic (T-O) phase tuners. The E-O phase shifter is used to modulate the microdisk's on-off state, providing high tuning efficiency and nanosecond-level switching speeds, while the two T-O phase tuners serve an auxiliary role to enable fine-tuning and compensation for fabrication variations.

*B. Optical Power Map of the E-O Switch*

Figure 5(a) shows the measured on-chip losses and channel crosstalk across various optical paths. The loss performance of 56 optical paths in this E-O switch was recorded, with measured crosstalk ranging from -42.8 to -51.9 dB and on-chip losses between 8.7 and 19.2 dB. The variation in loss is primarily attributed to the inter-layer couplers and fabrication deviations. In Figure 5(a), 8 out of 64 paths are unavailable due to fabrication defects and wire bonding damage. A tunable laser is used to launch a continuous-wave signal at C-band wavelengths, the FSR of this E-O switch is tested to be around 27.2 nm.

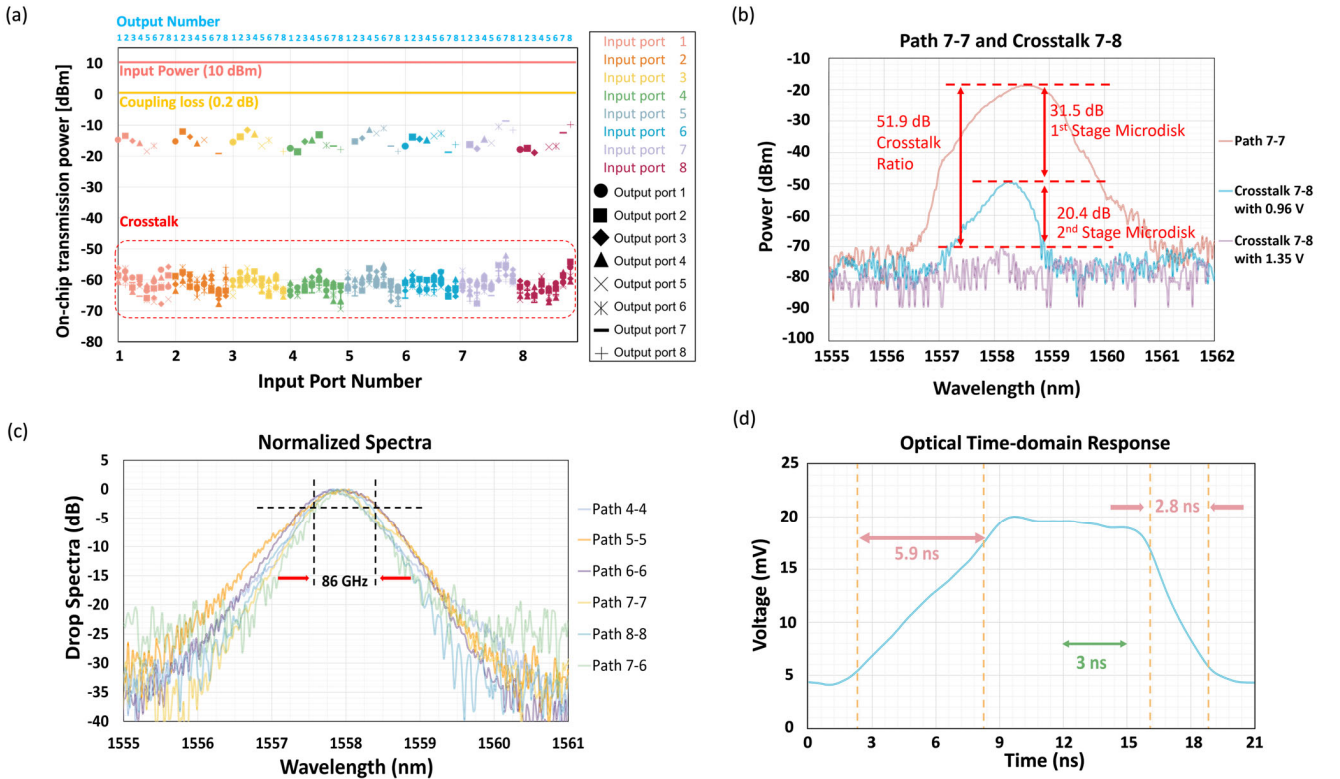

Fig. 5. (a) Measured optical power map of 8 × 8 microdisks based E-O switch. (b) Crosstalk ratio measurement: signal power at output port 7 and crosstalk at output port 8. (c) Normalized transmission spectra for some representative optical paths. (d) Optical time-domain response of E-O switch.

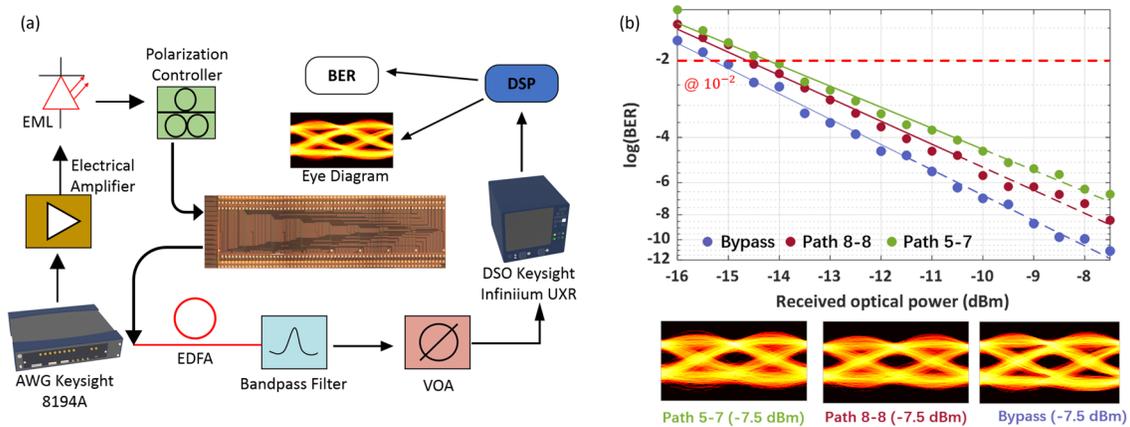

Fig. 6. (a) BER measurement setup. (b) BER of 25 Gb/s OOK transmission for path 8-8, 5-7, and bypass with eye diagrams.

*C. Extinction Ratio and Crosstalk of the E-O Switch*

To evaluate the crosstalk suppression performance, path 7-7 was chosen as an example route, and the optical power at output port 8 was measured to assess crosstalk leakage. The measured transmission spectra are shown in Figure 5(b), where the red curve presents the signal transmission of path 7-7, while the light blue and purple curves denote the channel responses when the 7th microdisk as output bus bank 8 is biased on and off, respectively. Based on Figure 5(b), the first- and second-order crosstalk ratios are 31.5 dB and 20.4 dB, respectively. The total crosstalk ratio of 51.9 dB is achieved. At the same time, all measured micro resonators feature an on-off extinction ratio exceeding 25.5 dB.

*D. Optical Passband, Switching Power, and Switching Time*

The normalized transmission spectra across different optical paths are presented in Figure 5(c), showing a 3 dB passband of this E-O switch of over 86GHz. The voltage range of the on-state bias for the micro-disks is 1.1–1.5 V, while the off-state bias voltage ranges from 0 to 0.28 V. The total power consumption per path is around 25 mW. To measure the switching time in the E-O switch, we employ a signal generator to produce a 50 MHz square wave with a 50% duty cycle and rise/fall times of 0.1ns. The time-domain response is shown in Figure 5(d). The measured switching rise and fall times (10%-90%) are 5.9 ns and 2.8 ns, respectively.

*E. BER and Eye Diagram Measurement*

Figure 6(a) details the BER measurement setup. The BER and eye diagrams are computed from the received measurements using offline generated transmitted signals. The signals are uploaded to an arbitrary waveform generator (AWG, Keysight M8194A) operating at 100 GSa/s for integer oversampling. Subsequently, the generated 25 Gb/s OOK signal is modulated by a low-cost C-band electro-absorption modulated laser (EML). The signal is then amplified by an erbium-doped fiber amplifier (EDFA) and filtered by a bandpass filter to reduce out-of-band noise. At the receiver side, the signal is sampled using a digital storage oscilloscope (DSO, Keysight Infiniium UXR) operating at 256 GSa/s.

Digital Signal Processing (DSP) is applied to the received signal, including frequency offset compensation and clock synchronization to align the transmitted and received signals. Figure 6(b) shows the experimental results, where paths 8-8 and 5-7 exhibit power penalties relative to the passive bypass case (no optical switch in the path) of 0.7 dB and 1.1 dB at BER with $10^{-2}$, respectively. It should be noted that the limited memory of AWG limits the transmission to a 100000-bit $2^{15}-1$ pseudo-random binary sequence (PRBS). The estimated BER for high received optical powers was therefore estimated by sweeping the decision thresholds [24]. The three lines (solid and dotted) in Figure 6(b) represent the fitted lines of the measured BER for different paths.

## V. DISCUSSION

The characterization of the T-O and E-O switches reveals that each features a distinct set of performances, in terms of switching time, on-chip loss, fabrication, packaging complexity, as well as the ease of control. Compared to the E-O switch, the T-O version offers lower on-chip loss and reduced system-wide complexity, making it well-suited for circuit-oriented applications. In contrast, the E-O device exhibits an unparalleled feature of nanosecond switching time with lower power consumption which gives it greater potential for latency-sensitive applications, such as high-performance computing systems, and time-sensitive network (TSN) applications.

The switch topology plays a vital role in the scalability and also largely determines its loss and crosstalk which could compromise the signal integrity. These challenges can be mitigated through the deployment of tailored architecture, as well as the incorporation of new photonic structures, such as the 3D waveguide routing platform. The proposed 3D topology implemented on the Si-SiN-SiN 3D platform demonstrates its great potential in lowering switch loss, crosstalk, and footprint. In this work, the coupling efficiency between different layers is not fully optimized, but for the first time, we demonstrate the power of combining such a 3D integration platform with E-O actuators. Such a platform holds great potential for a broader range of applications, such as optical computing, imaging, and sensing.

## VI. CONCLUSION

This paper presents two Si-SiN-SiN T-O and E-O optical switches based on the S&S topology, both demonstrating ultralow crosstalk. Each switch contains a total of 128 switching elements (MRRs or microdisks) to achieve 8×8 strictly non-blocking switching. The measured on-off extinction ratio for each MRR and microdisk exceeds 25 dB. Specifically, the T-O switch exhibits crosstalk ratios ranging from -38.9 dB to -50.8 dB, with on-chip losses between 2.1 dB and 10.5 dB across multiple paths, with switching rise and fall times of 17.6 μs and 0.42 μs. The E-O switch demonstrates ultralow crosstalk ranging from -42.8 dB to -51.9 dB, with on-chip losses as low as 8.7 dB and switching times of 5.9 ns and 2.8 ns. Additionally, the 25Gb/s OOK BER performance for the E-O switch shows error-free operation with a penalty ranging between 0.7 and 1.1 dB. These optical switches also show significant potential for future data centers and optical communication networks.

## ACKNOWLEDGMENTS

This work was supported in part by the U.S. ARPA-E under ENLITENED Grant DE-AR00084, the UKRI-EPSRC by project QUDOS (EP/T028475/1), and the European Union's Horizon Europe Research and Innovation Program project PUNCH (101070560) and project INSPIRE (101017088).

## Data Availability

The data that support this paper are available in the University of Cambridge Repository at https://doi-org.ezp.lib.cam.ac.uk/10.17863/CAM.XXXX.